\documentclass{article}
\usepackage[T1]{fontenc}
\usepackage[utf8]{inputenc}
\usepackage{ismir} % Remove the "submission" option for camera-ready version
\usepackage{amsmath,cite,url}
\usepackage{graphicx}
\usepackage{color}
\usepackage{amssymb}
\usepackage{bm}
\usepackage{booktabs}
\usepackage{pifont}
\usepackage{enumitem}
\usepackage{graphicx}

\usepackage{caption} 
\usepackage{subcaption}

\usepackage[protrusion=true, expansion=true, tracking=small]{microtype}

\usepackage{xcolor}
\usepackage{xparse}
\usepackage[normalem]{ulem}

\newcommand{\inlineSubsection}[2]{
  \refstepcounter{subsection} % Manually increment subsection counter
  \vspace{0.1cm}\noindent\textbf{\thesubsection\ #1}\label{#2}% Format subsection title with number and label
}

\title{AudioCodex: Streaming Autoencoders for Live Audio}
\multauthor
{Marco Pasini$^2$ \hspace{0.5cm} Javier Nistal$^1$ \hspace{0.5cm} Cyran Aouameur$^1$ \hspace{0.5cm} Stefan Lattner$^1$ \hspace{0.4cm} György Fazekas$^2$}
{
$^1$Sony Computer Science Laboratories, Paris, France\\
$^2$Queen Mary University of London, UK\\
{\tt m.pasini@qmul.ac.uk}
}

\def\authorname{M. Pasini, J. Nistal, C. Aouameur, S. Lattner, and G. Fazekas}

\usepackage[bookmarks=false,pdfauthor={\authorname},pdfsubject={\pdfsubject},hidelinks]{hyperref}
\begin{document}

\maketitle

\vspace{-0.2cm}
\begin{abstract}
\vspace{-0.3cm}
We introduce AudioCodex, a fully causal autoencoder for live generative audio applications that compresses stereo audio into compact continuous latent sequences at $\sim 10\,\mathrm{Hz}$. It combines a Mean Flow magnitude autoencoder with latent hypersphere projection, bounded-noise augmentation, representation alignment, and adversarial phase recovery to learn representations that support both efficient downstream generation and high-quality reconstruction. Through extensive ablations, we show that AudioCodex outperforms the state-of-the-art in terms of generation-reconstruction trade-off, whilst remaining strictly causal in both encoding and decoding. This work paves the way towards truly real-time interactive and streaming generative audio applications. Supplementary results and audio samples are available in the website.\footnote{\tiny\url{https://sonycslparis.github.io/audiocodex-companion}}

\end{abstract}

\vspace{-0.4cm}
\section{Introduction}\label{sec:introduction}
\vspace{-0.2cm}
Compact, structured latent representations of audio are central to both generative modeling~\cite{stable_audio_2,music2latent,diffariff} and music information retrieval~\cite{clap,repr_mert,mule}. Downstream models operating in such spaces, rather than directly on waveforms, become substantially more efficient during training and inference. 

In reconstruction-oriented applications, especially in generative modeling, autoencoders are the standard choice, typically learning deterministic mappings between audio signals and latent representations~\cite {soundstream, encodec, dac, stable_audio_open}. Yet existing audio autoencoders often expose a fundamental tension: representations that preserve detailed acoustic information can be difficult for generative models to learn~\cite{vitok,genae,rae,scale_rae}, whereas highly regularized or compressed latents may improve generative modeling at the cost of reconstruction fidelity~\cite{perceptuallynoiseaugmentation}. Stochastic generative decoders help relax this tension \cite{music2latent,music2latent2,codicodec} by decoupling reconstruction and audio quality: the latent carries global structure, while the decoder plausibly synthesizes missing fine detail. The objective is therefore not exact sample reconstruction, but a representation that is both generative-friendly and perceptually high-quality after decoding~\cite{tiltedseesaw}.

Recently, real-time audio applications are gaining traction---e.g., conversational speech~\cite{fullduplex,moshi}, musical accompaniment~\cite{diffariff,wu2025streaming,realtimeaccompanimentgen}, interactive synthesis~\cite{rave,latentgranularsynth,brave}---requiring these systems to remain strictly causal. While existing work addresses causal audio autoencoding for speech, these methods typically produce discrete tokens rather than continuous latents~\cite{soundstream, encodec, moshi, focalcodecstream, streamcodec}, rendering them less suitable for certain generative modeling frameworks~\cite{vae,gan,diffusion_original,ho_denoising_2020,song_score-based_2020}.

We introduce AudioCodex, a \emph{fully causal and continuous} audio autoencoder designed for real-time generative use. The model is trained as a generative autoencoder on log-magnitude spectrograms, with a separate adversarial phase estimator for waveform recovery. This decoupling facilitates high-quality reconstruction by removing the noisy phase components from the autoencoder training process. The magnitude decoder is trained as a Mean Flow model~\cite{mean_flow}, enabling single-step decoding and stable training compared to other one-step frameworks~\cite{consistencymodel}. To the best of our knowledge, this is the first Mean Flow-based autoencoder in any modality.

Our contributions are fivefold: (1) a fully causal autoencoder for general high sample rate stereo audio, designed for streaming to and from $\sim10$\,Hz compressed representations; (2) a magnitude-only formulation with separate adversarial phase estimation, trained without direct reconstruction objectives; (3) a sphere-normalized bottleneck with bounded latent noise augmentation~\cite{noiseaugmentation, perceptuallynoiseaugmentation, scalingdiffusionautoencoders, sphereencoder}; (4) Representation Alignment (REPA)~\cite{repa} with a frozen self-supervised audio encoder~\cite{usad}; and (5) a comprehensive analysis of the reconstruction-generation trade-off using both reconstruction and downstream-generative metrics.

\vspace{-0.3cm}
\section{Related Work}\label{sec:related}
\vspace{-0.2cm}
Neural audio autoencoders have largely followed two directions. Discrete codecs combine convolutional encoder--decoder architectures with residual vector quantization~\cite{vqvae,soundstream,encodec,dac}, yielding representations that are well suited to reconstruction-based language-modeling objectives such as next-token prediction~\cite{copet_simple_2023,agostinelli_musiclm_2023}. However, they are less directly compatible with generative frameworks commonly defined in continuous spaces~\cite{vae,gan,diffusion_original,ho_denoising_2020,song_score-based_2020}.
Inspired by works in vision~\cite{diffae,diffusionimagecompression,scalingdiffusionautoencoders}, continuous audio autoencoders address this limitation~\cite{music2latent,music2latent2,codicodec,stable_audio}. Yet they still face a reconstruction-generation trade-off: latents optimized for faithful reconstruction can be difficult to model downstream, whereas more regularized representations may compromise reconstruction fidelity~\cite{genae,vitok,tiltedseesaw,gigatok,rae}. AudioCodex is designed to improve this trade-off while remaining strictly causal, unlike prior continuous autoencoders for general audio, which typically rely on bidirectional processing and non-causal decoding~\cite{music2latent,music2latent2,stable_audio,stable_audio_2,stable_audio_open}. This makes it suitable for streaming generative applications, where causal autoencoders have so far been mostly speech-oriented and token-based~\cite{moshi,focalcodecstream,streamcodec}.
Finally, AudioCodex separates magnitude and phase modeling. Since wrapped phase induces an irregular geometry in complex STFTs~\cite{music2latent,music2latent2,codicodec}, and phase errors can interfere with magnitude prediction~\cite{mp_senet}, we encode only log-magnitude spectrograms and recover phase separately. This connects our design to neural vocoding~\cite{wavegan,kumar_melgan_2019,hifigan,yamamoto_parallel_2020} while allowing the magnitude autoencoder to be trained with a single flow-based objective.

\vspace{-0.3cm}
\section{Background}\label{sec:background}
\vspace{-0.2cm}
AudioCodex builds on improved Mean Flows~\cite{mean_flow,imf} for generative autoencoding, Representation Alignment (REPA)~\cite{repa} for semantic feature distillation into the compressed latents, noise augmentation for latent regularization, and the R3GAN adversarial framework~\cite{r3gan} for phase-estimator training.

\vspace{-0.1cm}
\inlineSubsection{Improved Mean Flow (iMF) \cite{imf}. }{subsec:background_imf} 
A Rectified Flow (RF)~\cite{rectified_flow} model learns to reach a data sample from noise by integrating a velocity field. It is trained with data-noise interpolations $x_t=(1-t)x+t\epsilon$ as input to approximate the conditional instantaneous velocity $v_c=\epsilon-x$. Mean Flow~\cite{mean_flow} generalizes this idea by modeling the \emph{average} velocity between two noise levels $r<t$, rather than only the instantaneous velocity used in standard RF~\cite{rectified_flow}:

\begin{equation}
  \small
  u(x_t,r,t)=\frac{1}{t-r}\int_r^t v(x_\tau,\tau)\,d\tau.
\end{equation}
We can write this as $(t-r)u=\int_r^t v\,d\tau$, and differentiating w.r.t. $t$ and rearranging the terms gives us the Mean Flow identity:
\begin{equation}
  \small
  u(x_t,r,t)=v(x_t,t)-(t-r)\frac{d}{dt}u(x_t,r,t).
\end{equation}
Mean Flow then trains $u_\theta$ by matching this identity with a stop-gradient target:
\begin{equation}
  \small
  \mathcal{L}_{\mathrm{MF}}=\mathbb{E}\left\|u_\theta(x_t,r,t)-\mathrm{sg}[u_{\mathrm{tgt}}]\right\|_2^2
\end{equation}
\begin{equation}
  \small
  u_{\mathrm{tgt}}=v-(t-r)\left(v\,\partial_z u_\theta+\partial_t u_\theta\right)
  \label{eq:mf}
\end{equation}
where $\frac{d}{dt}u_\theta$ is computed as a JVP. In practice, $v$ is replaced by the conditional velocity $v_c=\epsilon-x$.
iMF~\cite{imf} reformulates training as a regression on instantaneous velocity. By rearranging Eq.~\ref{eq:mf} we get the predicted instantaneous velocity:
\begin{equation}
    \small
    V_\theta(x_t,r,t)=u_\theta(x_t,r,t)+(t-r)\,\mathrm{sg}\!\left[\mathrm{JVP}(u_\theta;v_\theta)\right],
    \label{eq:imf}
\end{equation}
where $v_\theta$ is predicted from the same network at the boundary condition $r=t$. The iMF objective is then
\begin{equation}
  \small
  \mathcal{L}_{\mathrm{iMF}}=\mathbb{E}\left\|V_\theta(x_t,r,t)-v_c\right\|_2^2,
\end{equation}
This loss provides an empirically stable~\cite{imf} velocity-regression view of the Mean Flow objective.

\inlineSubsection{Representation Alignment (REPA)~\cite{repa}}{subsec:background_repa} augments the training objective by aligning intermediate model features with those of a frozen teacher network using a cosine similarity loss. 
This provides semantically informed gradients early in training and encourages the latent space to organize around the structure already captured by the teacher encoder. In AudioCodex, REPA guides the magnitude autoencoder toward more generative-friendly representations without significantly affecting reconstruction performance.

\inlineSubsection{Noise Augmentation~\cite{perceptuallynoiseaugmentation,scalingdiffusionautoencoders}}{subsec:background_noise}
perturbs latent representations during training to improve downstream generation. Rather than training only on clean latents $\boldsymbol{z}$, the decoder is trained on noisy interpolants,
\begin{equation}
  \small
  \boldsymbol{z}' = (1-t)\boldsymbol{z} + t\,\boldsymbol{n},
  \qquad
  \boldsymbol{n} \sim \mathcal{N}(0,I),
  \qquad
  t \sim \mathcal{U}(0,1).
\end{equation}
This regularization promotes perceptually meaningful latent hierarchies, with coarse structures capturing salient content and finer variations encoding high-frequency detail. The latent variance is fixed through layer normalization before interpolation, preventing the encoder from counteracting the noise by increasing its scale. Inspired by prior work that synchronizes latent and data noise levels~\cite{scalingdiffusionautoencoders}, we propose a novel bounded noise augmentation strategy that significantly improves generative performance at limited to no reconstruction cost.

\inlineSubsection{R3GAN~\cite{r3gan}}{subsec:background_r3gan} uses a relativistic pairing objective~\cite{jolicoeur2019relativistic} where the discriminator compares fake and real samples directly:
\begin{equation}
  \small
  \mathcal{L}_{\mathrm{rel}}(\theta,\psi)=
  \mathbb{E}_{z\sim p(z),\,x\sim p_{\mathcal D}}
  \Big[f\big(D_\psi(G_\theta(z))-D_\psi(x)\big)\Big].
\end{equation}
To keep training stable, R3GAN applies zero-centered penalties on both real and fake data:
\begin{equation}
  \small
  R_1(\psi)=\frac{\gamma}{2}\,\mathbb{E}_{x\sim p_{\mathcal D}}\!\left[\|\nabla_x D_\psi(x)\|_2^2\right]
\end{equation}
\begin{equation}
  \small
  R_2(\theta,\psi)=\frac{\gamma}{2}\,\mathbb{E}_{x\sim p_\theta}\!\left[\|\nabla_x D_\psi(x)\|_2^2\right]
\end{equation}
The discriminator maximizes $\mathcal{L}_{\mathrm{rel}}-R_1-R_2$, while the generator minimizes $\mathcal{L}_{\mathrm{rel}}$. The relativistic objective improves diversity and avoids mode collapse, and $R_1{+}R_2$ damps unstable gradients on both real and fake data manifolds.

\vspace{-0.2cm}
\section{AudioCodex}\label{sec:model}
\vspace{-0.1cm}
\begin{figure*}[t!]
    \centering
    \begin{subfigure}[l]{0.48\textwidth}
        \centering
        \includegraphics[height=2in]{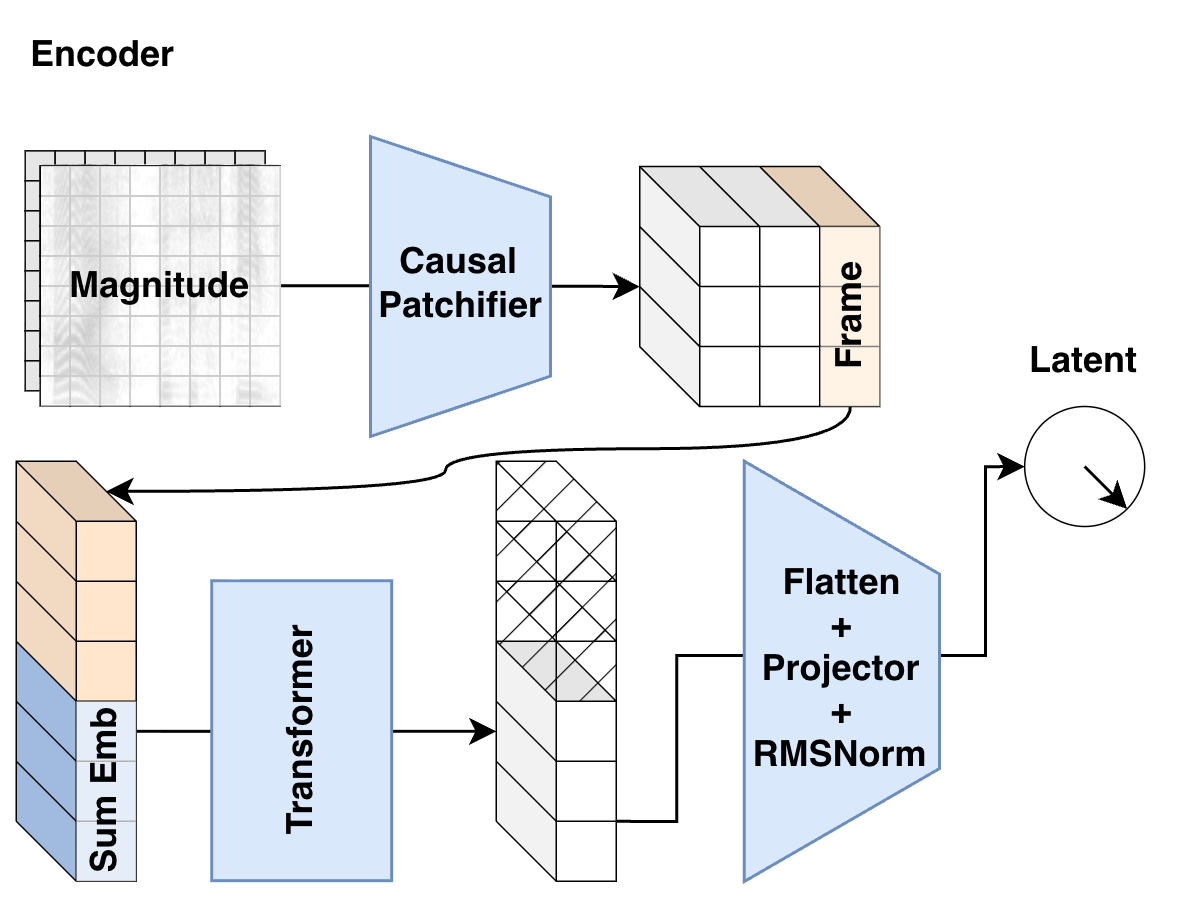}
        \caption{\footnotesize The causal encoder produces compact sphere-normalized latents. Learnable summary embeddings (\emph{Sum Emb}) are appended to each frame before the frequency-wise transformer.}

    \end{subfigure}%
    \hfill
    \begin{subfigure}[r]{0.48\textwidth}
        \centering
        \includegraphics[height=2in]{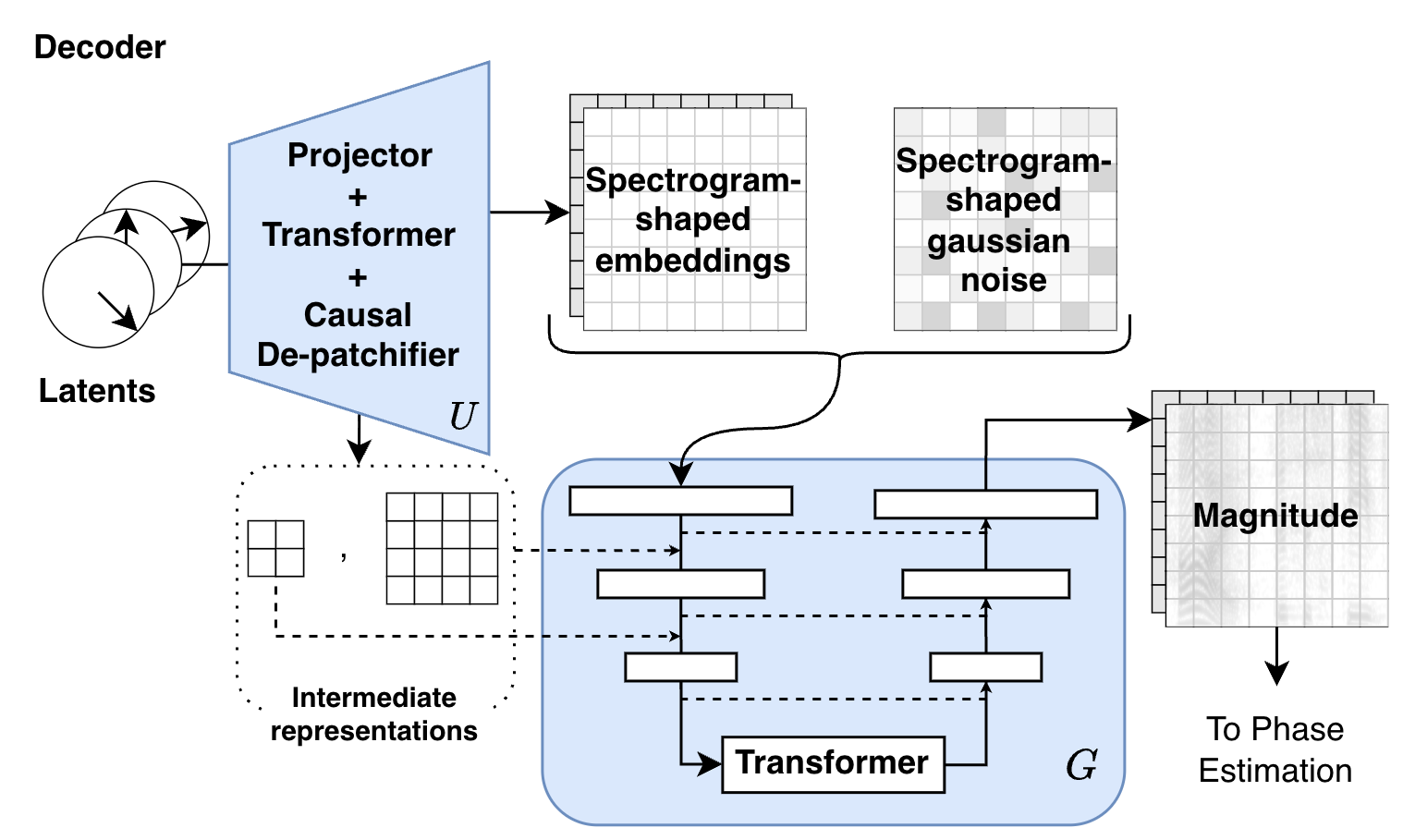}
        \caption{\footnotesize $G$ receives upsampled latent representations and spectrogram-shaped gaussian noise. $U$ provides input-level conditioning and multi-scale guidance to $G$ before magnitude-to-waveform reconstruction.}

    \end{subfigure}
    \vspace{-0.3cm}
    \caption{AudioCodex's autoencoder main architecture. Blue denotes learnable components.}
    \label{fig:audiocodex}
    \vspace{-0.4cm}

\end{figure*}

AudioCodex consists of two independently trained stages: a
\emph{magnitude autoencoder} that maps log-magnitude spectrograms to
and from a continuous latent representation, and a \emph{phase
estimator} that reconstructs phase from the decoded magnitudes.
This decoupling keeps the rapidly varying phase structure out of the
autoencoder objective, addressing the poor phase reconstructions often
observed in complex-STFT autoencoders~\cite{music2latent,
music2latent2,codicodec}. While this decoupling certainly adds complexity, our goal is to keep the training processes of the two modules simple and scalable.

\inlineSubsection{Audio representation.}{subsec:representation}
Given an input waveform $w$, we compute its Short-Time Fourier Transform (STFT) and retain only the log-magnitude $x = \bigl(20\log_{10}(\lvert \mathrm{STFT}(w)\rvert + \epsilon) - \mu\bigr)/\sigma$, where $\mu$ and $\sigma$ are scalar normalization parameters. Stereo audio is processed by treating the two channels as feature maps stacked along the channel axis. 

\inlineSubsection{Magnitude Autoencoder.}{subsec:architecture} 
Fig.~\ref{fig:audiocodex} depicts AudioCodex's magnitude autoencoder. 
In the encoder, a convolutional patchifier progressively downsamples the input spectrogram along time and frequency, resulting in a sequence of feature frames with temporal resolution $f_z$. A set of $K$ learnable summary embeddings is appended to each temporal frame~\cite{music2latent2,codicodec}. Each summary-augmented frame is then processed independently by a frequency-wise transformer. After the transformer, only the updated summary embeddings are retained, concatenated, flattened, and projected through the bottleneck to form a compact latent representation at resolution $f_z$.

The decoder consists of an upsampler $U$ and a generator $G$. The upsampler mirrors the encoder, lifting the compact latents into a spectrogram-shaped representation that matches the input resolution of $G$. This representation is concatenated with Gaussian noise at the generator input~\cite{codicodec}, while intermediate feature maps from $U$ are injected into $G$'s U-Net at matching resolution levels. In this way, $U$ provides both early conditioning and multi-scale guidance for the 1-step generation. The generator follows the same architectural principles as the encoder: a time-causal convolutional patchifier, a frequency-wise transformer, and a de-patchifier, augmented with standard U-Net skip connections.

Causality is enforced by using time-causal convolutions in all patchifiers and causal group normalization in all convolutional blocks, where normalization statistics are accumulated only from past context. Transformer layers use layer normalization and operate along the frequency axis independently for each $\sim10$\,Hz time frame. At inference time, the model can run incrementally in streaming mode by caching activations and processing only newly available input~\cite{rave,brave}.

\inlineSubsection{Bottleneck and Bounded Noise Augmentation}{subsec:bottleneck}. 
Sphere Encoder~\cite{sphereencoder} shows that projecting autoencoder latents onto the hypersphere and performing noise augmentation results in a latent distribution that is easy to sample from. We thus adopt the idea and RMS-normalize the encoder latents $\boldsymbol{z} \in \mathbb{R}^{d_z}$~\cite{rmsnorm}.
During training, we perform bounded noise augmentation. This matches the Sphere Encoder idea: noisy latent clouds are formed by adding Gaussian noise and projecting back to the sphere. Our implementation keeps the same intuition but uses interpolation rather than direct addition:
\begin{equation}
  \small
  \boldsymbol{\tilde z}_i = \mathrm{RMSNorm}\bigl((1-\alpha_i)\boldsymbol{\hat z}_i + \alpha_i\boldsymbol{\hat n}_i\bigr),
  \quad
  \alpha_i \sim \mathcal{U}(0, t_i),
\end{equation}
where $\boldsymbol{\hat n}_i$ is a sphere-normalized Gaussian noise vector and $t_i$ is the data-space noise level used for the corresponding noisy magnitude spectrogram in batch element $i$. Thus the latent corruption is capped per batch element: when the input is only mildly noisy, the latent stays mildly noisy too, which preserves access to latent high-frequency information. 
This can also be considered as a simplified form of noise synchronization~\cite{scalingdiffusionautoencoders}, without additional hyperparameters or layernorm. In practice, this preserves reconstruction performance, while still encouraging a generation-friendly representation~\cite{perceptuallynoiseaugmentation}.

\inlineSubsection{Training.}{subsec:training}
The magnitude autoencoder is trained using the iMF framework. For a normalized log-magnitude spectrogram $x$ and noise $\epsilon\sim\mathcal{N}(0,I)$, we define
\begin{equation}
  \small
  x_t=(1-t)x+t\epsilon,
  \qquad
  v_c=\epsilon-x.
\end{equation}
The start noise level $t$ is sampled from a logit-normal distribution, and the end time $r$ is sampled as
\begin{equation}
  \small
  r = \bigl(1-\xi^{\alpha}\bigr)t,\quad \xi\sim\mathcal{U}(0,1),
\end{equation}
The original Mean Flow formulation samples $t, r$ independently, enforces $r<t$ by swapping, and finally sets $r=t$ $75\%$ of the times. We instead first sample $t$ as in standard rectified-flow training, and only then sample $r$ conditionally from $t$. This avoids distorting the marginal distribution of $t$. Many noise levels where $r\approx t$ are still sampled (through the $\alpha$-exponentiated transform), so the model keeps a strong instantaneous-velocity learning signal and smoothly propagates it to larger jumps.
Given $(x_t,t,r)$, we first estimate the instantaneous velocity at $r=t$
\begin{equation}
  \small
  \hat v_\theta(x_t,t)=u_\theta(x_t,t,t),
\end{equation}
which is then used as the JVP tangent to evaluate $\frac{d}{dt}u_\theta$ at $(x_t,r,t)$. The final velocity prediction is
\begin{equation}
  \small
  V_\theta(x_t,r,t)=u_\theta(x_t,r,t)+(t-r)\,\mathrm{sg}\!\left[\frac{d}{dt}u_\theta(x_t,r,t)\right].
\end{equation}
We use the adaptive MSE loss proposed in Mean Flow, with an additional perceptual frequency weighting. Let $\Delta_i=V_{\theta,i}-v_{c,i}$ be the per-sample velocity error and $M$ a fixed mel-derived frequency weight. The implemented per-sample loss is
\begin{equation}
  \small
  \mathcal{L}_{\mathrm{MF}}=
  \frac{1}{B}\sum_i\frac{\,||M\odot \Delta_i||^2_2\,}
  {\left(\mathrm{sg}||\Delta_i||^2_2+c\right)^p},
  \quad
  M=\frac{1}{1+\frac{F}{700}}.
\end{equation}
where $p\in[0,1]$ controls sub-quadratic behaviour, $c>0$ is a numerical constant, stop-gradient is applied to the denominator term, and $F$ is the range of frequency bins of the STFT configuration in use.

Classifier-free guidance (CFG) is distilled into one-step training targets. Concretely, for $\sim10\%$ of batch elements during training, we run two predictions at $r=t$: one conditional and one unconditional,
\begin{equation}
  \small
  \hat v_i^{z}=V_\theta(z_{t,i},t_i,t_i\mid z),
  \quad
  \hat v_i^{\varnothing}=V_\theta(z_{t,i},t_i,t_i\mid \varnothing).
\end{equation}
We then form a distilled target velocity
\begin{equation}
  \small
  \tilde v_{c,i}=v_{c,i}+\mathbf{1}_{[t_{\min},t_{\max}]}(t_i)\,\omega\,\bigl(\hat v_i^{z}-\hat v_i^{\varnothing}\bigr),
\end{equation}
and train the main forward pass against $\tilde v_c$ on that subset. Outside the subset, $\tilde v_c=v_c$. Also note that we apply CFG only in the $(0.2, 0.8)$ noise level interval~\cite{intervalcfg}. The model is explicitly informed about mode through a learned conditioning embedding with three states (distilled-CFG mode, conditional mode, unconditional mode), injected through the adaptive normalization layers of all modules.

% \cyran{this paragraph about REPA feels like a repetition of 3.2, can be shortened ?} 
We further use Representation Alignment (REPA)~\cite{repa}. A frozen self-supervised teacher encoder produces a feature sequence $Y$. 
We extract one designated hidden representation from each of our three modules (encoder, upsampler, and decoder transformers), and project each sequence to the teacher representation shape with distinct 2-layer alignment probes. If $\hat Y^{(m)}$ denotes the projected feature sequence from each module $m$, the implemented REPA loss is:

\begin{equation}
  \small
  \mathcal{L}_{\mathrm{REPA}}=
  \frac{1}{3}\sum_m\Bigl(1-\cos\bigl(\hat Y^{(m)},Y\bigr)\Bigr),
\end{equation}
evaluated over all time positions and batch elements.
The final magnitude-autoencoder objective is

\begin{equation}
  \small
  \mathcal{L}_{\mathrm{AE}}=\mathcal{L}_{\mathrm{MF}}+\lambda_{\mathrm{REPA}}\mathcal{L}_{\mathrm{REPA}}
  \label{eq:repa}
\end{equation}

\inlineSubsection{Phase Estimator.}{subsec:phase}
The magnitude autoencoder does not model phase, so waveform reconstruction requires a separate, fully causal phase estimator. During training, the phase estimator takes ground-truth log-magnitude spectrograms 
% (with the same STFT parameters as the magnitude model) %cyran: obvious ? 
and Gaussian noise as input, and predicts phase information. To avoid phase discontinuities, we represent phase as interleaved cosine and sine channels: each frequency bin $k$ is represented as $[\cos(\phi_k), \sin(\phi_k)]$, normalized to unit magnitude during training.
The architecture mirrors the magnitude generator: a time-causal convolutional U-Net with the same temporal downsampling factor ($\sim10$\,Hz), but without transformer layers.
Training follows the R3GAN framework (Sec.~\ref{sec:background}) with two discriminators. The \emph{main discriminator} is magnitude-conditioned and evaluates whether a predicted phase matches a given magnitude.
We then reconstruct the waveform using the inverse STFT, and compute the power-compressed magnitude mel-spectrogram, possibly using a different STFT configuration and power-compression factor $\gamma$, and use it as the input to an \emph{auxiliary discriminator}~\cite{hifigan,bigvgan}.
Let $D_{\mathrm{main}}$ denote the main magnitude-conditioned discriminator and $D_{\mathrm{aux}}$ denote the auxiliary discriminator. The generator loss is
\begin{equation}
  \small
  \mathcal{L}_G = \mathcal{L}_{\mathrm{rel}}(D_{\mathrm{main}}) + \lambda_{\mathrm{aux}} \mathcal{L}_{\mathrm{rel}}(D_{\mathrm{aux}}),
\end{equation}
where $\mathcal{L}_{\mathrm{rel}}$ is the relativistic loss from R3GAN (Sec.~\ref{sec:background}). The discriminator loss is
\begin{equation}
  \small
  \mathcal{L}_D = \mathcal{L}_{\mathrm{rel}}(D_{\mathrm{main}}, D_{\mathrm{aux}}) + \gamma_{\mathrm{main}} R_{1+2, \mathrm{main}} + \gamma_{\mathrm{aux}} R_{1+2, \mathrm{aux}}.
\end{equation}
where the relativistic loss is evaluated jointly across both discriminators, and $R_1, R_2$ denote the zero-centered gradient penalties on real and fake data respectively, with fixed weights $\gamma_1$ and $\gamma_2$. 

\vspace{-0.3cm}
\section{Experimental Setup}\label{sec:experiments}
\vspace{-0.2cm}
\inlineSubsection{Training data.}{subsec:data}
We train AudioCodex on a mixed collection of permissively licensed datasets, using the same data recipe for both the magnitude autoencoder and the phase estimator. The training mixture is: Unsupervised People's Speech \cite{ups} (in-the-wild speech/vocal data that also includes music and general audio, 40\% sampling weight), Freesound \cite{freesound} (general audio, 30\%), and FMA \cite{fma} (music, 30\%). All audio samples are kept in their original sampling rate ($44.1$ or $48$\,kHz).

\inlineSubsection{Evaluation protocol.}{subsec:eval_protocol}
Our evaluation is organized into three parts: (i) targeted ablations, (ii) comparison to baselines, and (iii) streaming inference speed. The ablations are the core of this study and cover the main design choices of both the magnitude autoencoder and the phase estimator. Baseline comparison and latency benchmarking are then used to position the final selected configuration in terms of quality and real-time usability.

For evaluation, we build four domain-specific sets and collect 5k non-silent clips of 10\,s per domain: stems from MUSDB18-HQ \cite{musdb18_hq}, music from MusicCaps \cite{agostinelli_musiclm_2023}, speech from HiFiTTS \cite{hifitts}, and general audio from SoundBible \cite{soundbible}. Before cropping, all source audio is resampled to 44.1\,kHz when needed, and mono files are converted to stereo by channel duplication. Most source material is already stereo at either 44.1\,kHz or 48\,kHz. We release metadata for all evaluation samples, including the specific crops sampled, to ensure reproducibility.

\inlineSubsection{Metrics.}{subsec:metrics}
We use Fr\'echet Audio Distance (FAD)~\cite{fad} computed with both VGGish~\cite{vggish} and CLAP~\cite{clap} embeddings. For reconstruction experiments, to evaluate audio quality, we report \emph{reconstruction FAD} (rFAD), obtained by comparing embeddings from $10$\,s reconstructions with those from the original $10$\,s test samples. rFAD uses VGGish embeddings, while rFAD$_{\mathrm{C}}$ uses CLAP embeddings. When comparing to baselines, we also report L1 distances on dB-scale spectrograms and mel-spectrograms, respectively denoted \emph{STFT} and \emph{MEL}. We do not include SI-SDR~\cite{sisdr} since it measures exact-phase matching, and in our case phase is stochastically reconstructed.

To evaluate the quality of the latents for downstream generation, we use a fixed, lightweight proxy task: we first obtain latent sequences from a trained encoder, then train a small-scale latent rectified-flow transformer (4 layers, model dimension 512) for 40k iterations on latent sequences with a length of 16, and decode generated samples back to waveform space. FAD is then measured.
A length of 16 corresponds to $\sim1.5$\,s at our target latent rate ($\sim10$\,Hz, assuming 44.1\,kHz audio). For baseline representations produced at higher temporal rates (e.g., 20\,Hz), we concatenate adjacent timesteps along the channel axis so the transformer is still trained on length-16 sequences covering the same \(\sim1.5\,\)s span, which keeps the metric fair for different rates and compression ratios. For downstream generation experiments, we report \emph{generative FAD} (gFAD), obtained by comparing samples from the lightweight latent generator against the target data distribution after decoding. gFAD only uses VGGish embeddings since the $\sim1.5$\,s samples are too short for CLAP. 

\inlineSubsection{Ablation protocol.}{subsec:ablation_protocol}
All ablations for both the magnitude autoencoder and the phase estimator are trained for 200k iterations with fixed learning rate $5e-5$.
For the magnitude autoencoder, the reference configuration is: Mean Flow with data-space prediction ($x$-prediction), sphere-normalized bottleneck with $d_z=256$, bounded latent noise augmentation, CFG distillation weight 0.5, causal encoder/decoder, and $\alpha=2.5$. 
For the phase estimator, the selected reference configuration is: causal model, both discriminators enabled, power-compression factor $\gamma=0.4$, and $\lambda_{\mathrm{aux}}=1$. Non-targeted factors remain fixed for each ablation.
For ablations, we reduce the evaluation sets to 1k clips per domain and for brevity only report the average of the in-domain r/gFADs. This makes ablation metrics not directly comparable with Tab.~\ref{tab:autoencoder_main}.

\inlineSubsection{Baselines.}{subsec:baselines}
We compare against five representative autoencoders. M2L refers to a recent continuous audio autoencoder using consistency models~\cite{music2latent}; CDC denotes a hybrid continuous/discrete representation model~\cite{codicodec}, which we run in continuous mode; SAO denotes the VAE used in Stable Audio Open~\cite{stable_audio_open}; and DAC denotes a discrete residual-vector-quantized codec~\cite{dac}, which is only included as a reference since it is not directly comparable.
Our evaluation deliberately does not consider compression ratios: we are mainly interested in finding the best trade-off between reconstruction and generation, even at the cost of lower compression. 

\inlineSubsection{Implementation details.}{subsec:implementation}
AudioCodex processes stereo audio using the representation defined in Sec.~\ref{subsec:representation}. We use an STFT with window size $W=2048$ and hop size $H=1024$, and normalize log-magnitude spectrograms with $\mu=-35$ and $\sigma=75$.

For the magnitude autoencoder (Sec.~\ref{subsec:architecture}), the convolutional patchifiers use 4 resolution levels, with $64$ base channels and per-resolution channel multipliers $[1,2,4,8]$, $[3,3,3,1]$ layers per-level, and downsampling strides $[(2,2),(1,4),(2,2)]$ across (time, frequency) axes. Everything is inverted for all de-patchifiers. Two $3x3$ convolutions and a middle SiLU~\cite{swish} are used in the residual blocks.
The transformer backbones in the encoder, upsampler, and generator each use $12$ layers, a model dimension of $512$, a head dimension of $128$, and a feed-forward expansion factor of $4$ with SwiGLU activations. At the bottleneck, the encoder outputs $K=8$ summary embeddings. 
We jointly train different compression regimes by randomly sampling $d_z \in [128,256,512]$ for each element in the batch, and investigate $d_z$'s impact on the reconstruction-generation trade-off (see Tab.\,\ref{tab:ablations}).
For REPA, we use USAD-Small \cite{usad} as the frozen teacher with $\lambda_{\mathrm{REPA}}=0.1$ (Eq.~\ref{eq:repa}), extracting features from the 4th layer in the upsampler and the generator, and the 8th layer in the encoder. We choose it since it was trained with both general audio and speech signals.
The dimension of the alignment probes is $512$. iMF uses $p=0.5$, $t \sim LogitNormal(\mu=0, \sigma=1)$, and $r$ is sampled using $\alpha=2.5$  (Eq.~\ref{eq:imf}). We optimize the magnitude autoencoder with CRAdam~\cite{cradam, adam, adamw}, using peak learning rate $10^{-4}$, cosine decay to zero, $(\beta_1,\beta_2)=(0.9,0.999)$, weight decay 0.01, and 10k warm-up steps. Training runs for 500k iterations with a batch size of 72 on 3 A100 GPUs, and we maintain an exponential moving average with decay 0.9999. The full magnitude autoencoder contains $\sim150$\,M parameters. 

The phase estimator uses the same causal UNet as $G$, and does not use the transformer blocks, which results in a $\sim70$\,M parameter model. Both discriminators use $64$ base channels, 6 resolutions levels, $[1, 2, 4, 8, 8, 8]$ as the per-resolution channel multipliers, $[3, 3, 3, 3, 3, 3]$ layer per-level. The downsampling strides for the main discriminator are $[(2,2),(1,4),(2,2),(1,4),(2,2)]$ for the (time, freq) axes, and for the auxiliary one we use $[(2,2),(2,2),(2,2),(2,2),(2,2)]$. Leaky ReLUs are used in the residual blocks. For the auxiliary discriminator, we use a STFT window of 1024 and hop size of 256, with $108$ mel bins. We use $\lambda_{\mathrm{aux}}=1$, $\gamma_{\mathrm{main}}=50$, $\gamma_{\mathrm{aux}}=0.1$. We train for 200k iterations with a fixed learning rate of $10^{-4}$, using Adam \cite{adam} with $(\beta_1,\beta_2)=(0.5,0.9)$ for both generator and discriminator.

\vspace{-0.4cm}
\section{Results}\label{sec:results}
\vspace{-0.2cm}
We provide audio examples and supplementary results on all audio domains at this link\footnote{\tiny\href{https://sonycslparis.github.io/audiocodex-companion}{https://sonycslparis.github.io/audiocodex-companion}}.

\begin{table}[t]
\centering
\small
\setlength{\tabcolsep}{2pt}
\caption{Architecture ablations for: a) the autoencoder, evaluated using r/gFAD; b) the phase estimator using rFAD.}

\begin{minipage}[t]{0.64\linewidth}
\begin{tabular}{lcc}
\toprule
\textbf{a) Autoencoder} &  &  \\
\toprule

\vspace{-0.3cm}\\
\textbf{\vspace{-0.1cm}Formulation} & 
\textbf{rFAD} &
\textbf{gFAD} \\
\specialrule{0.06em}{0.25em}{0.25em}

CM$_{\mathrm{cSTFT}}$ & 2.17 & 4.81 \vspace{0.2cm}\\
$v$-MF$_{\mathrm{mSTFT}}^{\varnothing\mathrm{CFG}}$ & 0.83 & 3.63 \vspace{0.2cm}\\
$x$-MF$_{\mathrm{mSTFT}}^{\varnothing\mathrm{CFG}}$ & \textbf{0.82} & \textbf{3.61} \\

\vspace{-0.2cm}\\
\textbf{\vspace{-0.1cm}Bottleneck} & \multicolumn{2}{c}{} \\
\specialrule{0.06em}{0.25em}{0.25em}

tanh
& 0.82 & 5.14 \\
layernorm
 & 0.76 & 4.63 \\
sphere
& 0.77 & 4.57 \\
\hspace{0.2cm} + bounded noise
& \textbf{0.73} & 3.95 \\
\hspace{0.5cm} + REPA
& 0.79 & 3.45 \\
\hspace{0.2cm} + $n\sim \mathcal{U}_{0,1}$ + REPA
& 0.99 & \textbf{3.21} \\

\vspace{-0.3cm}\\
\textbf{\vspace{-0.1cm}\boldsymbol{$d_z$}} & \multicolumn{2}{c}{} \\
\specialrule{0.06em}{0.25em}{0.25em}

$64$
& 1.29 & \textbf{2.58} \\
$128$
& 0.90 & 2.60 \\
$256$
& 0.79 & 3.45 \\
$512$
& \textbf{0.61} & 4.64 \\

\vspace{-0.3cm}\\
\textbf{\vspace{-0.1cm}CFG weight} & \multicolumn{2}{c}{} \\
\specialrule{0.06em}{0.25em}{0.25em}

0.0
& 0.82 & 3.61 \\
0.5
& \textbf{0.79} & \textbf{3.45} \\
1.0
& 0.86 & \textbf{3.45} \\

\vspace{-0.3cm}\\
\textbf{\vspace{-0.1cm}\boldsymbol{$\alpha$}} & \multicolumn{2}{c}{} \\
\specialrule{0.06em}{0.25em}{0.25em}

1.5
& \textbf{0.76} & \textbf{3.34} \\
2.5
& 0.79 & 3.45 \\
3.5
& 0.84 & 3.51 \\

\vspace{-0.3cm}\\
\textbf{\vspace{-0.1cm}Causality} & \multicolumn{2}{c}{} \\
\specialrule{0.06em}{0.25em}{0.25em}

Causal
& 0.79 & 3.45 \\
Non-causal
& \textbf{0.65} & \textbf{3.25} \\
% Block-causal
% & 1.22 / 3.50 & 0.97 / 4.61 & 0.56 / 2.52 & 0.29 / 2.13 & 0.76 / 3.19 \\

\bottomrule
\end{tabular}
\end{minipage}
\hfill
\begin{minipage}[t]{0.34\linewidth}
\begin{tabular}{lc}

% \caption{Phase-estimator ablations. We report rFAD, and lower is better. All runs are trained for 200k steps at fixed learning rate $5e-5$. Values are rounded to two decimals.}
\toprule
\multicolumn{2}{l}{\textbf{b) Phase Estimator}}\\
\toprule

\vspace{-0.3cm}\\
\textbf{\vspace{-0.1cm} \boldsymbol{$D$} setup} &  \textbf{rFAD}\\
\specialrule{0.06em}{0.25em}{0.25em}

$D_{\text{ph}}$ & 0.41 \\
$D_{\text{aux}}$ & 0.31 \\
both & \textbf{0.25} \\

\vspace{-0.3cm}\\
\textbf{\vspace{-0.1cm}\boldsymbol{$\gamma$}} &   \\
\specialrule{0.06em}{0.25em}{0.25em}

$\gamma=0.2$ & 0.30 \\
$\gamma=0.4$ & \textbf{0.25} \\
$\gamma=0.7$ & 0.32 \\

\vspace{-0.3cm}\\
\textbf{\vspace{-0.1cm}\boldsymbol{$\lambda_{\mathrm{aux}}$}} &   \\
\specialrule{0.06em}{0.25em}{0.25em}

0.5 & 0.33 \\
1.0 & \textbf{0.25} \\
1.5 & 0.28 \\
2.5 & 0.27 \\

\vspace{-0.2cm}\\
\textbf{\vspace{-0.1cm}Causality} &   \\
\specialrule{0.06em}{0.25em}{0.25em}

causal & \textbf{0.25} \\
non-causal & 0.27 \\
% block-causal & 0.67 & 0.44 & 0.10 & 0.08 & 0.32 \\

\bottomrule
\end{tabular}
\end{minipage}
\label{tab:ablations}
\vspace{-0.3cm}
\end{table}

\inlineSubsection{Magnitude Autoencoder Ablations.}{subsec:results_mag_ablation}
Tab.~\ref{tab:ablations} reports rFAD and gFAD for all magnitude autoencoder ablations, averaged over our four audio domains. 
% We calculate these metrics on a 1k subset of our test sets, so these results are not directly comparable to the baseline eval results below.
We first train a consistency model using complex STFT representations, similarly to Music2Latent and CoDiCodec (we use the framework presented in \cite{codicodec} with our causal architecture). Results show how this variant underperforms our Mean Flow decoupled framework.
We also show how $x$-prediction \cite{jit} is slightly better than $v$-prediction. For these first ablations, we do not distill CFG, so that we can compare with the consistency-trained variant where CFG distillation is not directly possible. 
For bottleneck design, using bounded noise augmentation and REPA improves generative quality compared to tanh~\cite{musika,music2latent}, layernorm~\cite{perceptuallynoiseaugmentation, scalingdiffusionautoencoders}, and plain RMS-Norm, at a minor reconstruction cost. Uniform noise augmentation (where the noise level is always sampled from $\mathcal{U}(0,1)$) severely degrades reconstruction.
Bottleneck dimensionality shows a clear frontier: larger $d_z$ helps reconstruction, while smaller $d_z$ helps generation. We thus use multiple $d_z=[128,256,512]$ for our final model, so users can choose the trade-off that best suits their setting.
For CFG distillation, 0.5 gives the best trade-off. For the $r$ noise level sampling parameter, $\alpha=1.5$ is the strongest. However, we notice frequent loss spikes during training with such a low $\alpha$. We note that when $\alpha\rightarrow\infty$ our framework becomes equivalent to rectified flow (stable), while $\alpha=0$ corresponds to consistency training (generally unstable). We thus choose to use $\alpha=2.5$ in our final model, which results in smooth training. Finally, the non-causal variant performs better, showing that the causal constraint has a detrimental effect on both metrics.

\inlineSubsection{Phase Estimator Ablations.}{subsec:results_phase_ablation}
Tab.~\ref{tab:ablations} reports the rFAD for different ablations. The best results are obtained by using both main and auxiliary discriminators with power-compressed mel-magnitudes ($\gamma=0.4$), and an auxiliary discriminator weight of 1. Surprisingly, training a non-causal phase estimator does not improve our metrics.

\begin{table}[t]
\centering
\small
\setlength{\tabcolsep}{2pt}
\caption{Comparison to baselines. Metrics are defined in \ref{subsec:metrics}. For all metrics, lower is better.
% \javier{2 decimas is perhaps enough here, particularly for such "rough" metrics like FAD}
}
\label{tab:autoencoder_main}
% \resizebox{\textwidth}{!}{
\begin{tabular}{lccccc}
\toprule
\textbf{Model}      &
% \multicolumn{5}{c}{\textbf{General}} &
% \multicolumn{5}{c}{\textbf{Speech}} &
% \multicolumn{5}{c}{\textbf{Music}} &
% \multicolumn{5}{c}{\textbf{Stems}} &
% \multicolumn{5}{c}{\textbf{Avg.}} \\
% \cmidrule(lr){2-6}
% \cmidrule(lr){7-11}
% \cmidrule(lr){12-16}
% \cmidrule(lr){17-21}
% \cmidrule(lr){22-26}
% &
\textbf{rFAD} & \textbf{rFAD$_C$} & \textbf{gFAD} & \textbf{STFT} & \textbf{MEL} \\
\midrule

\vspace{-0.25cm}\\
\textbf{\vspace{-0.1cm}Codec Reference} & \multicolumn{5}{c}{} \\
\specialrule{0.06em}{0.25em}{0.25em}

DAC~\cite{dac}
& 0.601 & 0.044 & N/A & 0.094 & \textbf{0.057} \\

\vspace{-0.25cm}\\
\textbf{\vspace{-0.1cm}Baselines} & \multicolumn{5}{c}{} \\
\specialrule{0.06em}{0.25em}{0.25em}

M2L~\cite{music2latent}
& 0.943 & 0.233 & 5.241 & 0.132 & 0.090 \\

CDC~\cite{codicodec}
& \textbf{0.304} & 0.088 & 5.065 & 0.114 & 0.079 \\

SAO~\cite{stable_audio_open}
& 1.151 & 0.155 & 9.518 & 0.106 & 0.068 \\

\vspace{-0.2cm}\\
\textbf{\vspace{-0.1cm}AudioCodex} & \multicolumn{5}{c}{} \\
\specialrule{0.06em}{0.25em}{0.25em}

% $d_z=64$
% & 1.584 & 0.186 & 6.222 & 0.125 & 0.084
% & 1.553 & 0.114 & 2.325 & 0.135 & 0.088
% & 1.258 & 0.101 & 4.997 & 0.131 & 0.084
% & 0.597 & 0.128 & 2.385 & 0.116 & 0.075
% & 1.248 & 0.132 & 3.982 & 0.127 & 0.083 \\

$d_z=128$
& 0.773 & 0.098 & \textbf{3.696} & 0.106 & 0.071 \\

$d_z=256$
& 0.593 & 0.088 & 4.168 & 0.098 & 0.065 \\

$d_z=512$
& 0.544 & \textbf{0.083} & 6.259 & \textbf{0.092} & 0.061 \\

% \vspace{-0.2cm}\\
% \textbf{\vspace{-0.1cm}\footnotesize AudioCodex, 2-step} & \multicolumn{5}{c}{} \\
% \specialrule{0.06em}{0.25em}{0.25em}

% % $d_z=64$
% % & 1.475 & 0.176 & 6.227 & 0.123 & 0.083
% % & 1.487 & 0.112 & 2.260 & 0.134 & 0.088
% % & 1.224 & 0.100 & 4.896 & 0.129 & 0.083
% % & 0.576 & 0.119 & 2.460 & 0.114 & 0.074
% % & 1.190 & 0.127 & 3.961 & 0.125 & 0.082 \\

% $d_z=128$
% & 0.734 & 0.094 & 3.807 & 0.104 & 0.070 \\

% $d_z=256$
% & 0.547 & 0.085 & 4.282 & 0.096 & 0.064 \\

% $d_z=512$
% & 0.510 & \textbf{0.080} & 6.388 & \textbf{0.090} & \textbf{0.060} \\

\bottomrule
\end{tabular}
\end{table}

\inlineSubsection{Baseline Comparison.}{subsec:results_baselines}
In Tab.~\ref{tab:autoencoder_main}, the final AudioCodex configuration is compared to M2L, CDC, SAO, and DAC under the same eval protocol (Sec.~\ref{subsec:eval_protocol}). Results show that AudioCodex matches or exceeds almost all baseline results, while being fully causal and enabling streaming. We note that none of the presented baselines is causal, and to the best of our knowledge, no general-purpose streamable autoencoder producing continuous latents currently exists. We also acknowledge that AudioCodex supports multi-step decoding, although we do not observe meaningful improvements on our metrics when using multiple steps. 2-step results are included in our supplementary page.

\inlineSubsection{Inference speed.}{subsec:results_streaming}
In Tab.~\ref{tab:benchmark}, we evaluate the inference speed of AudioCodex on 48\,kHz stereo audio using \texttt{torch.compile} on a single A100 GPU (batch size 1), with timings averaged over 128 runs. At the target latent rate, which is about 10\,Hz for 44.1\,kHz audio and 12\,Hz for 48\,kHz audio (roughly 0.1\,s and 0.085\,s per frame, respectively), the full causal pipeline operates in real time, with combined decoding (upsampler, generator, and phase estimator) achieving about 2× faster-than-real-time performance per latent step. When processing full 10-second sequences (non-streaming mode), efficiency improves due to reduced overhead, with the full decoding pipeline exceeding 100× real-time.

% \begin{table}[t]
% \centering
% \caption{Inference latency and real-time factor measured on 48\,kHz stereo audio. Per-timestep results represent streaming performance at $\sim$12\,Hz latent rate. Values are computed using \texttt{torch.compile} with batch size 1, averaged over 128 repetitions. $\text{RTF}$ indicates how many times faster than real-time each component executes.}
% \begin{tabular}{lrr}
% \toprule
% \textbf{Component} & \textbf{Time (ms)} & \textbf{RTF} \\
% \midrule
% \multicolumn{3}{c}{\textit{Single latent streaming}} \\
% Encode & 6.80 & 12.56x \\
% Upsample & 6.85 & 12.47x \\
% Decode & 23.90 & 3.57x \\
% Phase estimator & 10.63 & 8.03x \\
% \midrule
% Combined decoding & 41.38 & 2.06x \\
% \midrule
% \multicolumn{3}{c}{\textit{10s sample}} \\
% Encode & 18.91 & 528.8x \\
% Upsample & 21.93 & 455.9x \\
% Decode & 35.71 & 280.0x \\
% Phase estimator & 30.01 & 333.2x \\
% \midrule
% Combined decoding & 87.64 & 114.1x \\
% \bottomrule
% \end{tabular}
% \label{tab:benchmark}
% \end{table}

\begin{table}[t]
\small
\centering
\caption{Inference latency and real-time factor (RTF) measured on a NVIDIA A100 GPU with 48\,kHz stereo audio.}
\begin{tabular}{lrr}
\toprule
\textbf{Component} & \textbf{Time (ms)} & \textbf{RTF} \\
\midrule
\multicolumn{3}{c}{\textit{Single latent streaming}} \\
Encoding & 6.80 & 12.56x \\
Combined decoding & 41.38 & 2.06x \\
\midrule
\multicolumn{3}{c}{\textit{10s sample}} \\
Encoding & 18.91 & 528.8x \\
Combined decoding & 87.64 & 114.1x \\
\bottomrule
\end{tabular}
\label{tab:benchmark}
\vspace{-0.4cm}
\end{table}

\vspace{-0.5cm}
\section{Conclusion}\label{sec:conclusion}
\vspace{-0.2cm}
We presented AudioCodex, a fully causal continuous autoencoder for live audio. It compresses stereo audio to $\sim10$\,Hz latents and reconstructs audio with a one-step magnitude decoder plus a separate causal phase estimator. In our ablations and baseline comparisons, AudioCodex shows a strong practical balance between reconstruction quality and downstream generative usability, while keeping both encoding and decoding streamable.
This study also has clear limitations. Non-causal decoding can still achieve better quality than causal decoding. Our gFAD proxy uses short ($\sim1.5$\,s) samples, so it does not fully measure long-range structure. Future work should improve causal decoding quality and extend downstream evaluation to longer horizons.

\section{Acknowledgements}
This work is supported by the EPSRC UKRI Centre for
Doctoral Training in Artificial Intelligence and Music (EP/S022694/1) and Sony Computer Science Laboratories Paris. Fazekas was supported by the Leverhulme Trust and the Royal Academy of Engineering under the RAEng / Leverhulme Trust Research Fellowships scheme.

% For BibTeX users:
\bibliography{ISMIRtemplate}

\end{document}